\documentclass[10pt,twocolumn,aps,pra,english,superscriptaddress,amsmath,amssymb,floatfix,longbibliography,nofootinbib]{revtex4-2}
\usepackage[T1]{fontenc}
\usepackage[utf8]{inputenc}
\usepackage{color}
\usepackage{babel}
\usepackage{amsmath}
\usepackage{graphicx}
\usepackage{float}
\usepackage{wasysym}
\usepackage{esint}
\PassOptionsToPackage{normalem}{ulem}
\usepackage{ulem}
\usepackage{enumitem}
\usepackage{amsmath}
\makeatletter
\usepackage{braket}
\usepackage{xcolor}

\usepackage{dcolumn}
\usepackage{soul}
\usepackage{amsfonts}
\usepackage{slashed}
\DeclareUnicodeCharacter{02BC}{'}
\makeatother
\usepackage[unicode=true,pdfusetitle,bookmarks=true,bookmarksnumbered=false,bookmarksopen=false, breaklinks=false,pdfborder={0 0 0},pdfborderstyle={},backref=false,colorlinks=true] {hyperref}
\begin{document}
\title{Optimal Thermalization under Indefinite Causal Order with Identical and Asymmetric Baths}

% Please use the appropriate macro for each type of information

\author{Neeraj Sharma}
%\thanks{}
\affiliation{Department of Physics, Indian Institute of Technology Jammu, Jammu 181221, India}
\author{Parveen Kumar}
%\email[]{parveen.kumar@iitjammu.ac.in}
%\thanks{}
\affiliation{Department of Physics, Indian Institute of Technology Jammu, Jammu 181221, India}


\date{\today}

\begin{abstract}
Indefinite causal order (ICO), in which the order of quantum operations is placed in a coherent superposition, has been demonstrated to enhance various information-processing tasks. Here, we investigate its impact on the thermodynamic processes generated by thermalizing quantum channels. We consider a two-level system interacting with two thermal baths under a quantum SWITCH, with the channel order controlled coherently by an ancillary qubit. We derive closed-form expressions for the effective inverse temperature $\beta_f$ of the postselected system state for both identical and distinct bath temperatures, and identify the control-qubit parameters that maximize heating or cooling. Our analysis reveals how the diagonal and coherent components of the control-qubit state contribute separately to the temperature shift, and how their interplay enables departures from the thermal response attainable under protocols with a definite causal order within the thermodynamic setting considered here. Bath asymmetry enhances these effects, while reduced purity of the control qubit state suppresses them. These results provide a systematic framework for assessing SWITCH-based thermalization in the setting of indefinite causal order, and identify control-qubit coherence as a tunable resource.
\end{abstract}

\maketitle

%---------------------------------------------------------
\section{Introduction}
\label{sec:intro}

Quantum theory allows operations whose causal order is not fixed but can be placed in a coherent superposition. This possibility of indefinite causal order, often studied within the quantum SWITCH formalism~\cite{PhysRevA.88.022318}, permits two quantum channels to act on a target system in an order that is controlled coherently by an ancillary qubit.  Indefinite causal order (ICO) has been shown to provide operational advantages in a range of quantum-information tasks, including computation~\cite{PhysRevA.88.022318,PhysRevA.86.040301,COLNAGHI20122940,PhysRevLett.113.250402,PhysRevA.90.010101,PhysRevLett.113.250402,PRXQuantum.2.010320,PRXQuantum.2.030335,Koudia2023,Kechrimparis_2024,ghosh2024designing,mondal2025path}, communication~\cite{PhysRevLett.120.120502,Chiribella_2021,e21101012,PhysRevA.92.052326,PhysRevLett.117.100502,PhysRevLett.120.060503}, cooling~\cite{PhysRevLett.125.070603,PhysRevResearch.4.L032029,Goldberg_2023}, work extraction~\cite{PhysRevA.102.032215,Simonov_2025,PhysRevA.105.032217}, sensing~\cite{frey2019indefinite,PhysRevLett.124.190503,PhysRevA.103.032615,PhysRevLett.130.070803}, and metrology~\cite{PhysRevLett.124.190503,frey2019indefinite,Goldberg2023}. Related control-induced interference effects have also been reported in qubit-assisted scattering processes~\cite{PhysRevB.99.045430}. Recent experiments~\cite{procopio2015experimental,PhysRevLett.121.090503,PhysRevLett.122.120504,PhysRevResearch.2.033292,PhysRevLett.124.030502,PhysRevLett.129.100603,PhysRevLett.131.060803,rozema2024experimental,PhysRevApplied.23.054075} have even demonstrated the use of indefinite causal orders on several experimental platforms. These developments naturally raise the question of whether, and in what ways, indefinite causal order can influence other fundamental quantum processes.

In a recent work, Felce and Vedral~\cite{PhysRevLett.125.070603} investigated whether quantum mechanical uncertainty in the causal order of operations can provide an advantage in thermodynamic tasks. They showed that placing two identical thermalizing channels in an indefinite causal order can drive a two-level system to a state whose effective temperature differs from that of the baths, and proposed a refrigeration cycle that exploits this mechanism. Subsequent studies~\cite{PhysRevResearch.4.L032029,Goldberg_2023} have explored related questions in cooling, work extraction, and thermodynamic resource theories, and have identified causal coherence as a nonclassical ingredient relevant to thermodynamic tasks. A key insight emerging from these works is that interference between different causal orders can alter the system dynamics in a manner that cannot be reproduced by any fixed sequential application of thermal channels.

Felce and Vedral’s protocol implements thermodynamic operations as quantum channels acting on a system, with the channel order controlled by an ancillary qubit. After the protocol, a measurement on the control qubit leaves the system in a thermal state. The control qubit was initialized and measured along specific directions on the Bloch sphere, and the dependence of the final system state on the initial state and measurement direction of the control qubit was not analyzed. We address this gap by generalizing their protocol to allow (i) an arbitrary initial state of the control qubit, (ii) an arbitrary measurement direction, and (iii) two distinct thermal baths instead of identical ones. Within this more general setting, we derive a closed-form expression for the effective inverse temperature $\beta_f$ of a two-level system, and determine the optimal parameters that maximize the cooling or heating of the system state. Our results provide a broader framework for assessing the potential usefulness of indefinite causal order in thermodynamic applications such as refrigeration. They also quantify how coherence and measurement of the control qubit act as tunable thermodynamic resources, and identify the conditions under which the quantum SWITCH yields thermal responses unattainable by protocols with a definite causal order for the class of thermalizing channels considered in this work.

It is also useful to investigate whether the thermodynamic behavior induced by the quantum SWITCH considered in this work can be reproduced by protocols employing a definite causal order, such as adaptive strategies described within the quantum comb framework~\cite{PhysRevLett.101.060401,PhysRevA.80.022339}. In the present setting, this comparison can be addressed analytically because the thermalizing channels considered here reset the system to the corresponding thermal state independently of the input state. This property implies that any protocol with a definite causal order yields a final system state determined solely by the thermal state of the channel applied last, thereby establishing a fixed-order benchmark for the thermodynamic task studied. This perspective helps clarify the role of coherent control of causal order in shaping the effective thermalization of the target qubit.

This paper is organized as follows. Section~\ref{sec:setup} introduces the setup and the underlying model. Section~\ref{sec:identical-baths} presents the analytic treatment and optimization of the ICO protocol with identical baths. Section~\ref{sec:different-baths} extends the analysis to the case of distinct baths. Section~\ref{sec:fixed-order-comparison} provides a comparison with protocols employing a definite causal order. Finally, Section~\ref{sec:discussion} summarizes our findings and discusses their implications.

%---------------------------------------------------------
\section{The setup}
\label{sec:setup}

We consider a two-level system that interacts with two thermal baths at inverse temperatures $\beta_{T1}$ and $\beta_{T2}$, together with an ancillary control qubit that determines the causal order of these interactions. Each interaction with a thermal bath is modeled as a completely positive trace-preserving (CPTP) map acting on the system state $\rho$. Any such map $\mathcal{E}$ can be written using the Kraus decomposition as~\cite{nielsen2010quantum,jacobs2014quantum}
\begin{equation}
    \mathcal{E}(\rho)=\sum_i K_i \rho K_i^\dagger,
\end{equation}
where the Kraus operators $\{K_i\}$ satisfy $\sum_i K_i^\dagger K_i=\mathbb{I}$. The explicit Kraus operators corresponding to the thermalizing channels will be introduced in the following sections.

The system Hamiltonian is defined as $H_s=\Delta |1\rangle\langle 1|$, so the energy eigenstates are the ground state $|0\rangle$ and the excited state $|1\rangle$. For simplicity, we take the system to be initially in a thermal state,
\begin{equation}
\label{system-initial-state}
    \rho_i =\frac{1}{1+e^{-\beta_i \Delta}} \begin{pmatrix}
        1 & 0 \\
        0 & e^{-\beta_i \Delta}
    \end{pmatrix},
\end{equation}
where $\beta_i$ is the inverse temperature of the initial state.

The control qubit is initialized in an arbitrary state on the Bloch sphere,
\begin{equation}
    \rho_c = \frac{1}{2}\Big(\mathbb{I} + \,\vec{n}\cdot\vec{\sigma}\Big),
\end{equation}
where the Bloch vector is
\begin{equation}
\vec{n} = \big(r\sin\theta\cos\phi,\, r\sin\theta\sin\phi,\, r\cos\theta\big),
\end{equation}
with $r\in[0,1]$ denoting the magnitude of the Bloch vector, $(\theta,\phi)$ the polar and azimuthal angles, and $\vec{\sigma}=(\sigma_x,\sigma_y,\sigma_z)$ the Pauli vector.

\begin{figure}[H]
    \centering
    \includegraphics[width=1.0\columnwidth]{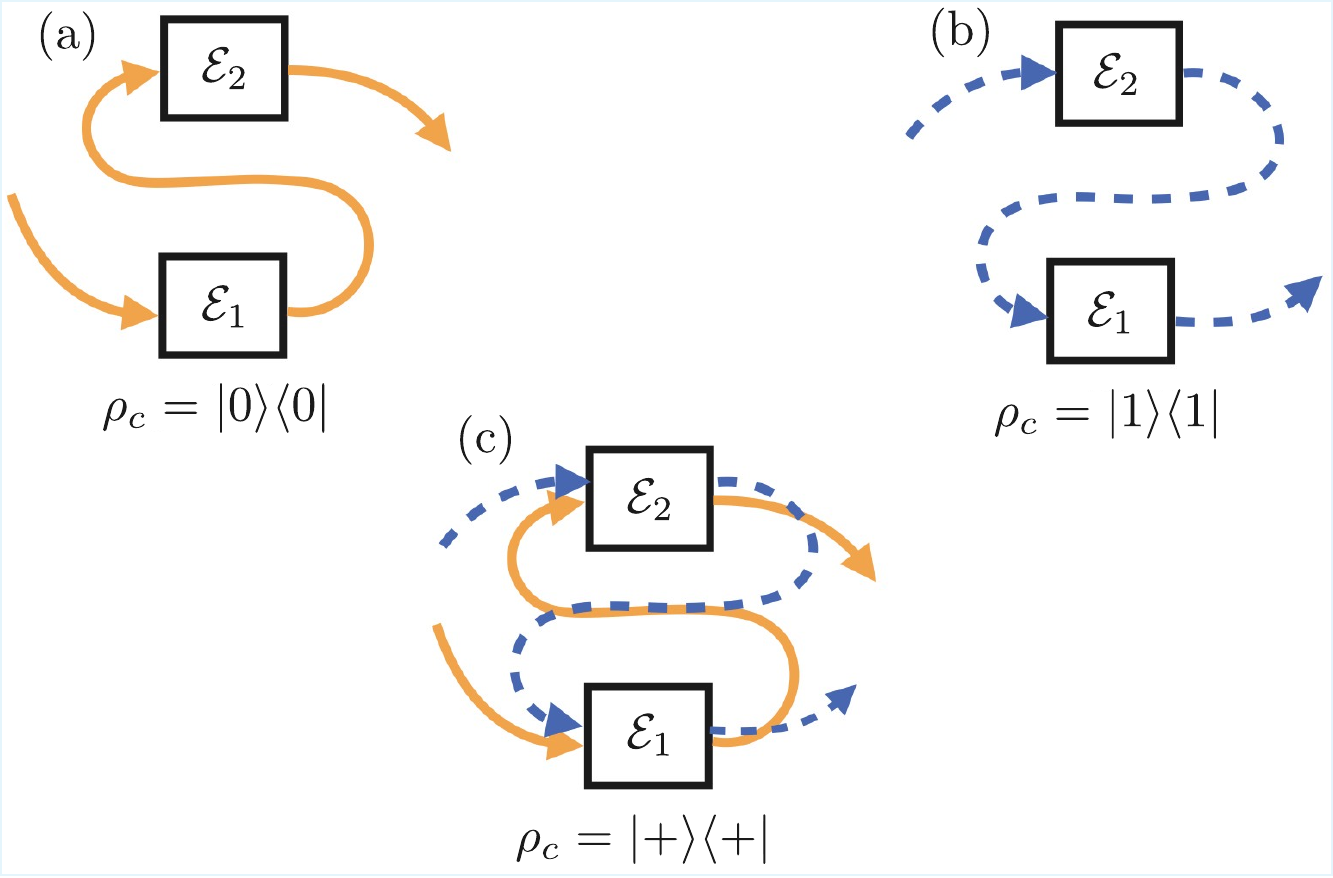} 
    \caption{Schematic of the ICO thermalization protocol. A control qubit, prepared in an initial state $\rho_c$, coherently determines whether the system undergoes the sequence of thermal channels $\mathcal{E}_1 \circ\mathcal{E}_2$ or $\mathcal{E}_2 \circ\mathcal{E}_1$. After the joint evolution, the control is measured in a given basis, and the system is postselected on the chosen outcome. The resulting system state $\rho_f$ is diagonal in the energy basis and can be described by an effective inverse temperature $\beta_f$, which may differ from the temperatures of the baths.}
    \label{fig:schematic}
\end{figure}

In a definite (direct) causal order setting, the thermal channels act in a fixed sequence: the system undergoes $\mathcal{E}_{1}$ followed by $\mathcal{E}_{2}$, or the order is reversed. In the quantum SWITCH, the order of the channels is placed in a coherent superposition, with the control qubit determining whether the system experiences $\mathcal{E}_{1}\circ\mathcal{E}_{2}$ or $\mathcal{E}_{2}\circ\mathcal{E}_{1}$. A schematic of the protocol is shown in Fig.~\ref{fig:schematic}. The SWITCH implements a joint CPTP map, $\mathcal{S}(\mathcal{E}_1,\mathcal{E}_2)$, on the combined system–control space. For two CPTP channels $\mathcal{E}_{1}$ and $\mathcal{E}_{2}$, the Kraus operators of the SWITCH are
\begin{equation}
\label{Kraus-op-quant-SWITCH}
    M_{ij}=|0\rangle\langle 0|_c ~\otimes~K_i^{(2)}K_j^{(1)} + |1\rangle\langle 1|_c~\otimes~K_j^{(1)}K_i^{(2)},
\end{equation}
where the subscript $c$ denotes the control qubit and $K_j^{(1)}$, $K_i^{(2)}$ are the Kraus operators of $\mathcal{E}_{1}$ and $\mathcal{E}_{2}$, respectively.
If the control qubit is in the state $\ket{0}$, the system encounters $\mathcal{E}_{1}$ before $\mathcal{E}_{2}$; if it is in $\ket{1}$, the order is reversed. The operators ${M_{ij}}$ act on the joint state according to
\begin{equation}
\label{joint-evoln}
\mathcal{S}(\rho_c\otimes\rho)
= \sum_{i,j} M_{ij}(\rho_c\otimes\rho) M_{ij}^\dagger,
\end{equation}
and the map $\mathcal{S}\equiv \mathcal{S}(\mathcal{E}_{1},\mathcal{E}_{2})$ returns the joint system–control state after the SWITCH.

We then measure the control qubit in a direction specified by the Bloch-sphere angles $(\Theta,\Phi)$. The measurement has two possible outcomes, corresponding to projections along $(\Theta,\Phi)$ and $(\pi-\Theta,\pi+\Phi)$, and we postselect the outcome associated with the direction $(\Theta,\Phi)$. This postselection on the control-qubit outcome induces a backaction on the system state. By construction, the resulting conditional system state remains diagonal in the energy eigenbasis of $H_s$ and can therefore be written as an effective thermal state with inverse temperature $\beta_f$. The dependence of $\beta_f$ on the initial control parameters $(r,\theta,\phi)$, the measurement direction $(\Theta,\Phi)$, and the bath temperatures $\beta_{T1}$ and $\beta_{T2}$ is the central focus of the analysis that follows.

\section{Indefinite causal order with identical baths}
\label{sec:identical-baths}

In this section, we examine the ICO protocol when the two thermal baths are identical and thermalize the system to the same temperature $T$. We first derive the effective final temperature attained by the system under the SWITCH, and then determine the control-qubit parameters that maximize the achievable cooling or heating.

\subsection{Final effective temperature of the system}
\label{subsec:effective-temp-identical-baths}

We begin with the case in which both thermal baths have the same inverse temperature $\beta_T$ and therefore drive the system toward the same thermal state. A convenient Kraus representation of such a thermalizing channel is~\cite{PhysRevLett.125.070603}
\begin{eqnarray}
    K_1 &=& \tfrac{1}{\sqrt{2}}A, \quad 
    K_2 = \tfrac{1}{\sqrt{2}}A\sigma_x, \nonumber\\
    K_3 &=& \tfrac{1}{\sqrt{2}}A\sigma_y, \quad 
    K_4 = \tfrac{1}{\sqrt{2}}A\sigma_z,
    \label{kraus-operators-same-channels}
\end{eqnarray}
where \(A=\sqrt{\rho^{(T)}}\) and
\begin{equation}
\label{the-matrix-M}
\rho^{(T)}=\frac{1}{1+e^{-\beta_T\Delta}}
\begin{pmatrix}
1 & 0\\[2pt]
0 & e^{-\beta_T\Delta}
\end{pmatrix}.
\end{equation}
When both channels correspond to $\beta_T$, the SWITCH acting on an initial product state $\rho_c\otimes\rho_i$ produces the joint state
\begin{eqnarray}
\mathcal{S}(\rho_c\!\otimes\!\rho_i)
&=&\sum_{i,j} M_{ij}(\rho_c\!\otimes\!\rho_i)M_{ij}^\dagger \nonumber\\
&=&\Big[\tfrac{1+r\cos\theta}{2}|0\rangle\langle0|_c
+\tfrac{1-r\cos\theta}{2}|1\rangle\langle1|_c\Big]\!\otimes\!\rho^{(T)} \nonumber\\
&+&\tfrac{r\sin\theta}{2}\Big[e^{-i\phi}|0\rangle\langle1|_c
+e^{i\phi}|1\rangle\langle0|_c\Big]\!\otimes\!\rho^{(T)}\rho_i \rho^{(T)}.
\label{joint-evolution-with-Mij-1}
\end{eqnarray}
The joint state is in general entangled between the system and the control; measuring the control in the computational basis $\{|0\rangle_c,|1\rangle_c\}$ recovers the ordinary thermalization to temperature $T$.

If the control qubit is instead projected onto an arbitrary Bloch direction,
\begin{equation}
|\psi_c^m\rangle
=\cos\tfrac{\Theta}{2}|0\rangle_c
+e^{i\Phi}\sin\tfrac{\Theta}{2}|1\rangle_c,
\end{equation}
the postselected (unnormalized) system state is
\begin{eqnarray}
\rho_f &\propto& \langle\psi_c^m|\mathcal{S}(\rho_c\!\otimes\!\rho_i)|\psi_c^m\rangle, \nonumber \\
&=&\frac{1}{2}\!\left[(1+r\cos\Theta\cos\theta)\rho^{(T)} \right. \nonumber \\
&\quad& + \left. r\sin\Theta\sin\theta\cos(\Phi-\phi)\,\rho^{(T)}\rho_i \rho^{(T)}\right],
\label{system-state-after-meas-T-unnormalized}
\end{eqnarray}
so that, after normalization, the conditional state of the system $\rho_f$ depends, in general, on the system's initial state \(\rho_i\), control qubit's purity, and the angles $\left(\theta,\Theta,\phi,\Phi\right)$ parametrizing control qubit's initial state and its measurement axis on the Bloch sphere. As before, the system's final state $\rho_f$ remains diagonal in the energy eigenbasis of $H_s$ and can be written as an effective thermal state,
\begin{equation}
\rho_f=\frac{1}{1+e^{-\beta_f\Delta}}
\begin{pmatrix}
1 & 0\\[2pt]
0 & e^{-\beta_f\Delta}
\end{pmatrix},
\label{rhof-defn}
\end{equation}
with \(\beta_f\) given by
\begin{widetext}
\begin{align}
\beta_f = \beta_T 
-\frac{1}{\Delta}
\ln\!\Bigg[
\frac{(1+e^{-\beta_i\Delta})(1+e^{-\beta_T\Delta})(1+r\cos\Theta\cos\theta)
+r e^{-(\beta_i+\beta_T)\Delta}\sin\Theta\sin\theta\cos(\Phi-\phi)}
{(1+e^{-\beta_i\Delta})(1+e^{-\beta_T\Delta})(1+r\cos\Theta\cos\theta)
+r\sin\Theta\sin\theta\cos(\Phi-\phi)}
\Bigg].
\label{beta-f-T}
\end{align}
\end{widetext}

Equation~(\ref{beta-f-T}) shows that two distinct features of the control qubit determine how the final inverse temperature deviates from the bath temperature. The factor $\left(1+r\cos\Theta\cos\theta\right)$ arises from the diagonal components of the control qubit in the measurement basis and modifies the relative weighting of the thermal contributions in the numerator and denominator. By contrast, the term $r\sin\Theta\sin\theta\cos\left(\Phi-\phi\right)$ originates from the coherences of the control qubit; it is therefore the contribution associated with interference between the two causal orders and vanishes whenever the control state is effectively classical.

Whether the system is cooled $\left(\beta_f > \beta_T\right)$ or heated $\left(\beta_f < \beta_T\right)$ is determined by the sign of the logarithm in Eq.~(\ref{beta-f-T}). If the numerator is smaller than the denominator, the logarithm is negative and $\beta_f>\beta_T$; if the numerator is larger, the opposite holds. Two limiting cases provide useful checks: for $r=0$ the coherence term vanishes and $\beta_f=\beta_T$, recovering ordinary thermalization, while for a pure control qubit $\left(r=1\right)$ the ICO-induced deviation is maximal for fixed angles and temperatures.

\subsection{Optimal control parameters for cooling and heating}
\label{subsec:optimization-identical-baths}

For a fixed initial state of the control qubit, i.e., for given \((r,\theta,\phi)\), the final inverse temperature \(\beta_f\) can be extremized by an appropriate choice of the measurement direction \((\Theta,\Phi)\). Analytically solving \(\partial_{\Theta}\beta_f=\partial_{\Phi}\beta_f=0\) yields two stationary points,
\begin{eqnarray}
\Theta &=& \cos^{-1}\!\left[r\cos(\pi-\theta)\right], \qquad 
\Phi = \phi, 
\label{beta-f-maxima}
\end{eqnarray}
and
\begin{eqnarray}
\Theta &=& \cos^{-1}\!\left[r\cos(\pi-\theta)\right], \qquad 
\Phi = \phi\pm\pi,
\label{beta-f-minima}
\end{eqnarray}
which correspond respectively to the maximum and minimum values of \(\beta_f\).

The azimuthal conditions \(\Phi=\phi\) and \(\Phi=\phi\pm\pi\) correspond to constructive and destructive interference between the two causal orders. Aligning the measurement phase with the control phase (\(\Phi=\phi\)) fixes the coherence term in one sign, while an antialigned phase (\(\Phi=\phi\pm\pi\)) reverses its sign and produces the opposite extremum.

The optimal polar angle \(\Theta=\cos^{-1}[\,r\cos(\pi-\theta)\,]\) captures the balance between the purity and orientation of the control qubit. For a pure control state (\(r=1\)), the measurement axis aligns to fully use the coherence set by \(\theta\). As the control becomes mixed (\(r<1\)), the optimal \(\Theta\) shifts toward the equator, consistent with the reduced coherence of the control state. Two simple limits are instructive. If the control lies on the equator (\(\theta=\pi/2\)), then \(\Theta_{\mathrm{opt}}=\pi/2\) for any \(r\). In contrast, when the control is polarized along \(z\) (\(\theta=0\)), the coherence term in Eq.~(\ref{beta-f-T}) vanishes for all \((\Theta,\Phi)\), so \(\beta_f=\beta_T\) independently of the measurement direction and ICO reduces to ordinary thermalization.

For \(r=0\), Eq.~(\ref{beta-f-T}) reduces to \(\beta_f=\beta_T\), recovering ordinary thermalization, while for \(r=1\) the deviation \(|\beta_f-\beta_T|\) is maximized at the optimal angles identified above. In the main text, we therefore focus on the pure-control case \(r=1\), which yields the largest ICO-induced temperature shifts and illustrates the full range of behavior. Results for mixed controls \(r<1\), which show the same qualitative behavior with smaller magnitude, are provided in the Appendix. 

The optimized cooling and heating points should be interpreted together with the
corresponding postselection probability. For a measurement along the Bloch
direction \((\Theta,\Phi)\), the success probability obtained from
Eq.~(\ref{system-state-after-meas-T-unnormalized}) takes the form
\begin{eqnarray}
p(\Theta,\Phi)
&=& \frac{1}{2}\Big[1
+r\cos\theta\cos\Theta  \nonumber \\
&+& r\,\gamma\,\sin\theta\sin\Theta\cos(\Phi-\phi)\Big],
\label{success-prob-general}
\end{eqnarray}
where
\begin{equation}
\gamma =
\frac{1+e^{-(\beta_i + 2\beta_T)\Delta}}
{\left(1+e^{-\beta_i \Delta}\right)\left(1+e^{-\beta_T \Delta}\right)^2}
\label{gamma-identical-baths}
\end{equation}
is a temperature-dependent coefficient satisfying \(0<\gamma<1\).  
Evaluating Eq.~(\ref{success-prob-general}) at the angles that extremize
\(\beta_f\) gives, for a pure control state \((r=1)\),
\begin{equation}
p_{\mathrm{opt}}^{(\pm)}
=\frac{1}{2}\sin^{2}\theta\,\big(1\pm\gamma\big),
\label{success-prob-opt}
\end{equation}
where the plus and minus signs correspond to the maximum and minimum values of
\(\beta_f\), respectively. Because \(0<\gamma<1\), at least one of the two
extremal branches necessarily has a reduced success probability. In the
low-temperature limit \(\beta_i\Delta\gg1\) and \(\beta_T\Delta\gg1\), one has
\(\gamma\simeq1\), in which case the less probable branch occurs only in a small
fraction of postselected runs. Thus, the extremal values of \(\beta_f\) are
accompanied by an operational trade-off between the achievable temperature shift
and the likelihood of obtaining the corresponding postselected outcome.

\section{Indefinite causal order with different baths}
\label{sec:different-baths}

We now consider the more general situation in which the two thermal baths have different inverse temperatures, \(\beta_{T1}\) and \(\beta_{T2}\). Unlike the identical-bath case, the two channels now drive the system toward distinct thermal states, and the interference between the two causal orders leads to a more intricate dependence of the final temperature on the control settings. In this section, we derive the resulting effective temperature of the system state and determine the optimal measurement parameters of the control qubit.

\subsection{Final effective temperature of the system}
\label{subsec:effective-temp-diff-baths}

Here, the system interacts with two distinct thermalizing channels, \(\mathcal{E}_1\) and \(\mathcal{E}_2\), and the control qubit places their order in a coherent superposition. Postselection on a control outcome yields a conditional system state that can be written as an effective thermal state with inverse temperature \(\beta_f\), which now depends on both bath temperatures and the complete set of control parameters.

The joint evolution of the system and control qubit is governed by the CPTP map \(\mathcal{S}\) introduced in Eq.~\eqref{joint-evoln}. Projecting the control qubit onto the Bloch-sphere direction \((\Theta,\Phi)\) yields the unnormalized conditional system state
\begin{align}
\rho_f 
&\propto\,
\langle\psi_c^m|\mathcal{S}(\rho_c\!\otimes\!\rho_i)|\psi_c^m\rangle \nonumber\\
&= \frac{1}{4}\Big[
(1-r\cos\theta)(1-\cos\Theta)\,\rho^{(T1)} \nonumber\\
&\quad +(1+r\cos\theta)(1+\cos\Theta)\,\rho^{(T2)} \nonumber\\
&\quad +\,2r\sin\Theta\sin\theta\,
\cos(\Phi-\phi)\,\rho^{(T1)}\rho_i \rho^{(T2)}
\Big],
\label{system-state-after-meas-unnormalized-two-different-thermal-channels}
\end{align}
where $\rho^{(T1)}$ and $\rho^{(T2)}$ are thermal states with temperature $T1$ and $T2$, respectively.

The normalized conditional state is diagonal in the energy basis, so it can be written as a thermal state with inverse temperature \(\beta_f\), and one obtains
\begin{widetext}
\begin{align}
\beta_f = -\frac{1}{\Delta}
\ln\!\Bigg[
\frac{(1+e^{-\beta_i\Delta})
\big[\alpha_1(1+r\cos\Theta\cos\theta)
-\alpha_2(\cos\Theta+r\cos\theta)\big]
+2\alpha_3 r\sin\Theta\sin\theta
\cos(\Phi-\phi)}
{(1+e^{-\beta_i\Delta})
\big[\alpha_4(1+r\cos\Theta\cos\theta)
+\alpha_2(\cos\Theta+r\cos\theta)\big]
+2r\sin\Theta\sin\theta
\cos(\Phi-\phi)}
\Bigg],
\label{beta-f-two-different-thermal-channels}
\end{align}
where
\begin{eqnarray}
\alpha_1 &=& 2e^{-(\beta_{T1}+\beta_{T2})\Delta}
+e^{-\beta_{T1}\Delta}+e^{-\beta_{T2}\Delta}, \label{alpha1}\\
\alpha_2 &=& e^{-\beta_{T1}\Delta}-e^{-\beta_{T2}\Delta}, \label{alpha2}\\
\alpha_3 &=& e^{-(\beta_i+\beta_{T1}+\beta_{T2})\Delta}, \label{alpha3}\\
\alpha_4 &=& 2+e^{-\beta_{T1}\Delta}+e^{-\beta_{T2}\Delta}. \label{alpha4}
\end{eqnarray}
\end{widetext}

Equation~\eqref{beta-f-two-different-thermal-channels} generalizes Eq.~\eqref{beta-f-T} to the case of different baths. Its structure parallels the identical-bath expression: the terms proportional to $\left(1+r\cos\Theta\cos\theta\right)$ originate from the diagonal components of the control state and determine the baseline weighting of the two thermal channels, while the term proportional to \(r\sin\Theta\sin\theta\cos(\Phi-\phi)\) arises from the control-qubit coherence and encodes the interference between the two causal orders. In the present case, the coefficients $\alpha_1,\alpha_2,\alpha_3$ and $\alpha_4$ introduce an additional dependence on \(\beta_{T1}\) and \(\beta_{T2}\), so bath asymmetry modifies how these diagonal and coherence contributions combine in the numerator and denominator of Eq.~\eqref{beta-f-two-different-thermal-channels}.

Several limiting observations follow directly from the expression. When the bath temperatures are equal $(\beta_{T1}=\beta_{T2})$, Eq.~\eqref{beta-f-two-different-thermal-channels} smoothly reduces to Eq.~\eqref{beta-f-T}.  
If one bath is significantly hotter or colder than the other, the corresponding exponential factors in $\alpha_1,\alpha_2,\alpha_3$, and $\alpha_4$ can dominate the prefactors in the numerator and denominator, and the resulting \(\beta_f\) reflects this imbalance together with any coherence-dependent contribution. Finally, as in the identical-bath case, setting \(r=0\) removes all coherence terms and reduces the expression to a purely classical combination of the two thermalizing operations.

These features complete the characterization of the final effective temperature in the asymmetric-bath setting.  
The next subsection examines the choice of control parameters that maximize cooling or heating for given bath temperatures.

\begin{figure}
\centering
\includegraphics[width=1.0\columnwidth]{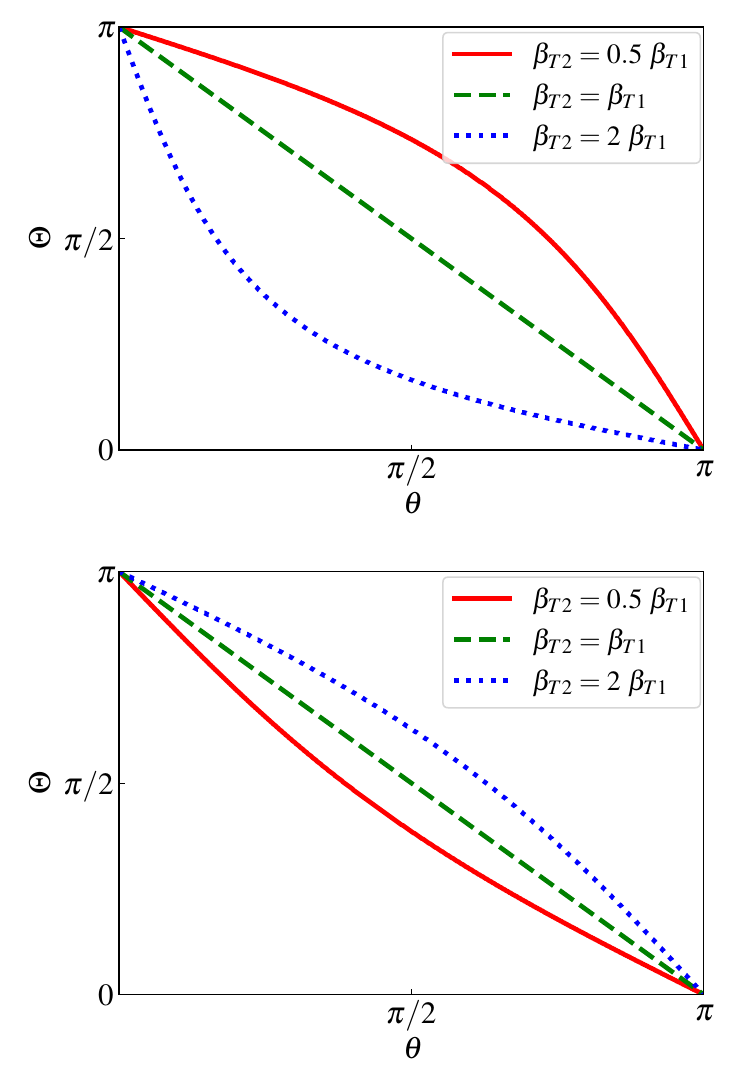}
\caption{
\label{fig:theta_m_vs_theta_c}
Optimal measurement polar angle $\Theta$ as a function of the initial control angle $\theta$ for a pure control qubit $(r=1)$. The three curves correspond to bath asymmetry parameters $n=\beta_{T2}/\beta_{T1}=0.5,~1.0,~2.0$. The optimal values identify the measurement directions that maximize or minimize the final inverse temperature.}
\end{figure}

\subsection{Optimal control parameters for cooling and heating}
\label{subsec:optimization-diff-baths}

\begin{figure*}
\centering
\includegraphics[width=1.0\textwidth]{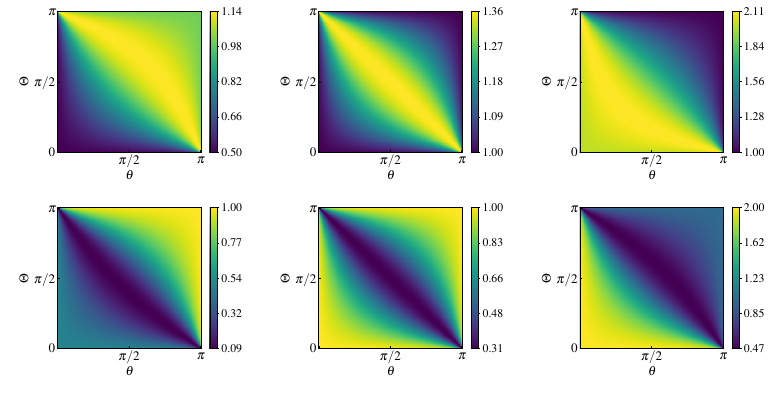}
\caption{Heat maps of the temperature shift $\Delta\beta(\Theta,\theta)=\beta_f-\beta_{T1}$ for optimal phase differences $\Delta\Phi=\Phi-\phi$. Top row: $\Delta\Phi=0$ (phase aligned, constructive interference); bottom row: $\Delta\Phi=\pi$ (phase anti-aligned, destructive interference). 
Columns (left to right) correspond to bath-asymmetry ratios $n=\beta_{T2}/\beta_{T1}=0.5,\;1.0,\;2.0$. 
For all panels, we use $\beta_{T1}\Delta=1.0$, $\beta_i=\beta_{T1}$, and the control qubit in a pure state $(r=1)$. Color indicates the deviation of the ICO output inverse temperature from the reference bath temperature.}
\label{fig-3:dbeta_vs_theta_m_vs_theta_c}
\end{figure*}

Analytic optimization of Eq.~\eqref{beta-f-two-different-thermal-channels} with respect to \((\Theta,\Phi)\) is challenging in the general asymmetric case. We therefore determine the extrema numerically. To parametrize the asymmetry, we write \(\beta_{T2}=n\,\beta_{T1}\). In all simulations, the system is initially assumed to be in a thermal state at temperature \(T1 \). The numerical optimization shows that \(\beta_f\) is maximized at \(\Phi=\phi\) and minimized at \(\Phi=\phi\pm\pi\). The optimal measurement angle $\Theta$ as a function of the initial polar angle $\theta$ of the pure control qubit, for the asymmetry parameters $n=0.5,~1.0,~2.0$, is shown in Fig.~\ref{fig:theta_m_vs_theta_c}. The dependence of the temperature shift \(\Delta\beta(\Theta,\theta)=\beta_f-\beta_{T1}\) on \((\Theta,\theta)\) at these optimal azimuthal settings is displayed in Fig.~\ref{fig-3:dbeta_vs_theta_m_vs_theta_c}, which presents heat maps of \(\Delta\beta\) for asymmetry parameter \(n=0.5\), \(1.0\), and \(2.0\). The upper and lower rows correspond to phase alignment \(\Delta\Phi=\Phi-\phi=0\) and anti-alignment \(\Delta\Phi=\Phi-\phi=\pi\), respectively, while the three columns show different values of \(n\). The figure displays the case of a pure control state (\(r=1\)); analogous plots for a mixed control state are provided in the Appendix. 

Figure~\ref{fig-4:extrema_vs_n} summarizes the globally optimized extremal temperatures as a function of the bath-asymmetry parameter 
$n=\beta_{T2}/\beta_{T1}$. For a pure control qubit $(r=1)$, the ICO protocol enables both enhanced cooling $(\beta_f^{\textrm{max}}>\beta_{T1})$ and enhanced heating $(\beta_f^{\textrm{min}}<\beta_{T1})$, with the achievable range growing as the asymmetry between the baths increases. This reflects the fact that a larger temperature contrast provides more room for interference between the two causal orders to modify the effective thermal response. For mixed control states (dashed curves), the reduced coherence suppresses these deviations, but the overall dependence on $n$ remains similar. These trends illustrate how bath asymmetry and control-qubit coherence jointly determine the operational advantage obtained in the quantum SWITCH protocol over protocols with a definite causal order.

\begin{figure}[t]
\centering
\includegraphics[width=1.0\columnwidth]{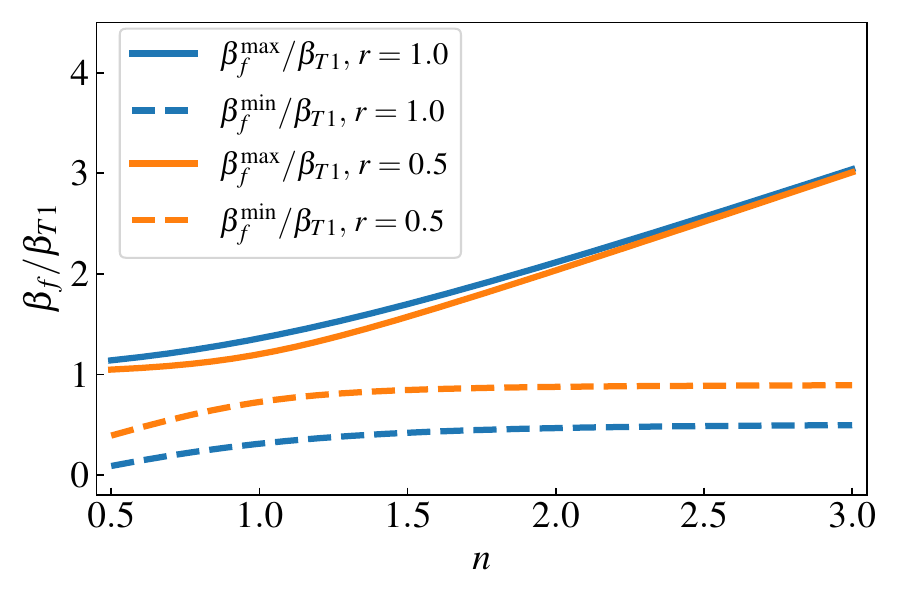}
\caption{Globally optimized output inverse temperatures $\beta_f^{\max}$ and $\beta_f^{\min}$, normalized to $\beta_{T1}$, as functions of bath-asymmetry parameter $n=\beta_{T2}/\beta_{T1}$. Solid lines show results for a pure control qubit $(r=1.0)$, and dashed lines correspond to a control qubit initialized in a mixed state $(r=0.5)$. For this figure, we use $\beta_{T1}\Delta=1.0$ and $\beta_i=\beta_{T1}$.}
\label{fig-4:extrema_vs_n}
\end{figure}

\section{Comparison with definite causal order}
\label{sec:fixed-order-comparison}

Before turning to the broader implications of the above results, it is instructive to examine how the thermodynamic behavior obtained under the quantum SWITCH compares with that arising in protocols employing a definite causal order. The most general adaptive protocol with a definite causal order can involve preparing the target qubit together with an ancillary system, allowing joint operations between successive uses of the channels, and conditioning the final system state on a measurement outcome of the ancilla. Such strategies are commonly described within the framework of quantum combs~\cite{PhysRevLett.101.060401,PhysRevA.80.022339}. 
The importance of comparing SWITCH-based protocols with adaptive strategies of definite causal order has also been emphasized in other contexts, where it was shown that definite-order schemes can outperform the quantum SWITCH for certain metrological tasks~\cite{PhysRevA.109.062435}.

In the present thermodynamic setting, each thermalizing map acts on the same target qubit and is used only once. Consequently, any admissible definite-order protocol relevant to the present thermodynamic task can be reduced, without loss of generality, to a sequential application of the two channels, possibly interspersed with joint system–ancilla operations. A parallel use would correspond to enlarging the task to multiple target subsystems and is therefore outside the scope of the present setting.

The structure of the thermalizing channels considered here allows the fixed-order behavior to be analyzed explicitly. From Eq.~(\ref{kraus-operators-same-channels}), the action of a thermalizing channel associated with bath temperature $T$ on an arbitrary system operator $X$ is
\begin{equation}
\mathcal{E}_T(X)=\frac{1}{2}A\left(X+\sigma_x X\sigma_x+\sigma_y X\sigma_y+\sigma_z X\sigma_z\right)A,
\end{equation}
where $A=\sqrt{\rho^{(T)}}$. Using the identity
\begin{equation}
X+\sigma_x X\sigma_x+\sigma_y X\sigma_y+\sigma_z X\sigma_z
=2\,\mathrm{Tr}(X)\,I,
\end{equation}
valid for any $2\times2$ operator $X$, one obtains
\begin{equation}
\mathcal{E}_T(X)=\rho^{(T)}\,\mathrm{Tr}(X).
\label{eq:thermal_reset}
\end{equation}
Thus, each thermalizing channel resets the target qubit to the corresponding thermal state independently of the input. This input-independent action is specific to the thermal channels considered here and enables a direct comparison between fixed-order and ICO protocols.

After the first thermalizing channel, the joint state of the system and ancilla may undergo an arbitrary CPTP map and therefore need not remain a product state before the second channel acts. Consider now an arbitrary joint density matrix of the system and ancilla immediately before the second channel in a definite-order protocol. Writing it in block form with respect to an orthonormal ancilla basis $\{|0\rangle,|1\rangle\}$,
\begin{equation}
\rho_{SA}=
\begin{pmatrix}
X_{00} & X_{01} \\
X_{10} & X_{11}
\end{pmatrix},
\end{equation}
where each block $X_{mn}$ is an operator acting on the system Hilbert space. Equivalently,
\begin{equation}
\rho_{SA}=\sum_{m,n=0}^{1} X_{mn}\otimes |m\rangle\langle n|.
\end{equation}
Application of the second thermalizing channel on the system alone yields
\begin{equation}
(\mathcal{E}_{T2}\otimes I_A)(\rho_{SA})=
\begin{pmatrix}
\mathcal{E}_{T2}(X_{00}) & \mathcal{E}_{T2}(X_{01}) \\
\mathcal{E}_{T2}(X_{10}) & \mathcal{E}_{T2}(X_{11})
\end{pmatrix}.
\end{equation}
Using Eq.~(\ref{eq:thermal_reset}), this becomes
\begin{equation}
(\mathcal{E}_{T2}\otimes I_A)(\rho_{SA})
=
\rho^{(T2)}
\otimes
\begin{pmatrix}
\mathrm{Tr}(X_{00}) & \mathrm{Tr}(X_{01}) \\
\mathrm{Tr}(X_{10}) & \mathrm{Tr}(X_{11})
\end{pmatrix}.
\end{equation}
The matrix in the second factor is precisely the reduced state of the ancilla, $\rho_A=\mathrm{Tr}_S(\rho_{SA})$. Hence the joint state factorizes after the action of the second channel,
\begin{equation}
(\mathcal{E}_{T2}\otimes I_A)(\rho_{SA})
=
\rho^{(T2)}\otimes \rho_A.
\end{equation}
Any subsequent operation on or postselection of the ancilla therefore leaves the conditioned system state unchanged, namely equal to the thermal state of the bath applied last. An analogous conclusion holds when the order of the channels is reversed. Convex mixtures of the two fixed-orders can thus generate only mixtures of the corresponding bath thermal states.

This establishes the fixed-order reference behavior relevant for the present thermodynamic task. In particular, no protocol with a definite causal order—irrespective of the use of ancillas, adaptive operations, or postselection—can produce a target state that differs from the thermal state of the last-applied bath. Convex mixtures of definite-order strategies therefore generate only mixtures of the corresponding bath thermal states. By contrast, the postselected quantum SWITCH protocol yields target states characterized by effective inverse temperatures that can differ from those associated with either bath. These deviations originate from coherence between causal orders and cannot be reproduced within any fixed-order framework using the same resources.

\section{Discussion}
\label{sec:discussion}

Our results demonstrate how the quantum SWITCH modifies thermalization when a two-level system interacts with two thermal baths under coherently controlled channel order. We derived closed-form expressions for the effective inverse temperature $\beta_f$ of the postselected system state, identified the control qubit settings that maximize cooling or heating, and analyzed how these effects depend on the bath temperatures and the purity of the control qubit.

The behavior of $\beta_f$ can be traced to two complementary contributions: the diagonal part of the control qubit state, which rescales thermal weights, and the coherence term, which encodes interference between the two causal orders. The interplay of these two contributions determines the deviation from fixed-order thermalization, with the coherence term—absent in any definite-order protocol—providing the mechanism responsible for this behavior. Bath asymmetry amplifies these deviations by increasing the contrast between the two thermalizing channels, while the reduced purity of the control qubit state suppresses them by diminishing the coherence available to generate interference. 

Our work generalizes the setting introduced by Felce and Vedral in 
Ref.~\cite{PhysRevLett.125.070603}, who considered identical baths and fixed preparation and measurement directions of the control qubit. The present analysis shows how the thermal response depends on the full set of control parameters and how bath asymmetry reshapes the ICO-induced shift in the effective temperature. These results place earlier observations of ICO-assisted cooling within a broader framework and delineate the mechanisms responsible for the observed temperature shifts. In particular, the comparison with fixed-order strategies clarifies that the observed temperature shifts originate from interference enabled by quantum coherence between the two causal orders rather than from adaptive processing alone.

Several limitations should be noted. The temperature shifts we study arise in the postselected branch of the SWITCH, and a complete thermodynamic assessment would require accounting for the associated success probability and the energetic cost of measurement. Similarly, the present work focuses on effective temperatures of diagonal states; a more detailed thermodynamic accounting would examine heat flows and entropy changes along the full joint evolution. Finally, robustness to noise in the control qubit and to imperfections in the thermalizing channels remains to be explored.

Even with these caveats, the results indicate that quantum coherence in the control qubit serves as a tunable resource for modifying thermalization. Possible extensions include analyzing ICO-assisted work extraction, designing thermodynamic cycles that exploit coherence between causal orders, and identifying broader classes of quantum channels for which quantum SWITCH protocols yield advantages beyond those attainable with adaptive protocols of definite causal order. Recent works have highlighted the distinction between advantages of specific implementations, such as the quantum SWITCH, and those attributable more generally to indefinite causal order~\cite{zhao2025quantum}. In this context, it would be interesting to clarify under what conditions such advantages can be attributed specifically to indefinite causal order.

\begin{acknowledgments}
We acknowledge fruitful discussions with Kyrylo Snizhko. NS acknowledges funding from the IIT Jammu Institute Fellowship. PK acknowledges support from the IIT Jammu Initiation Grant No. SGT-100106
\end{acknowledgments}

\appendix

\section{ICO with Control qubit initialized in a mixed state}

\begin{figure*}[t]
\centering
\includegraphics[width=1.0\textwidth]{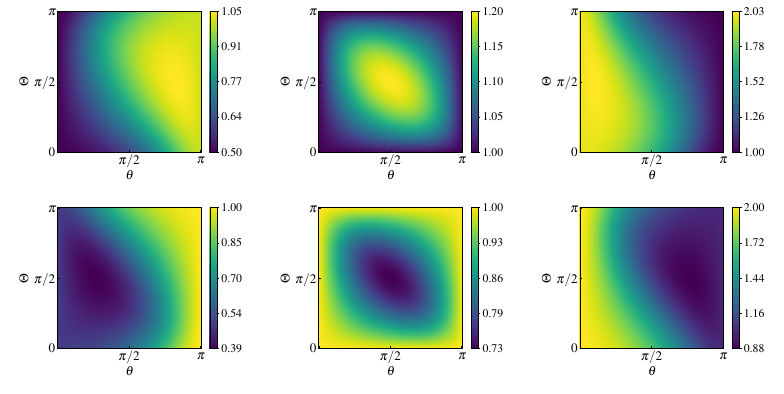}
\caption{Heat maps of $\Delta\beta(\Theta,\theta)=\beta_f-\beta_{T1}$ for the case of control qubit initialized in the mixed state with the length of the Bloch vector $r=0.5$. Layout and parameter choices are identical to Fig.~\ref{fig-3:dbeta_vs_theta_m_vs_theta_c} from the main text. The reduced purity suppresses coherence-driven contributions and narrows the range of temperatures attainable through ICO.}
\label{fig-5:dbeta_vs_theta_m_vs_theta_c}
\end{figure*}

For completeness, Fig.~\ref{fig-5:dbeta_vs_theta_m_vs_theta_c} shows the analog of Fig.~\ref{fig-3:dbeta_vs_theta_m_vs_theta_c} when the control qubit is prepared in a state with Bloch-vector magnitude $r = 0.5$. The qualitative structure of the temperature landscape remains unchanged: constructive and destructive interference between causal orders creates regions of enhanced cooling and heating.

A reduced purity of the control qubit state suppresses the coherence-dependent contributions in Eqs.~(\ref{beta-f-T}) and (\ref{beta-f-two-different-thermal-channels}), thereby narrowing the range of temperatures that can be reached through ICO. As a result, the temperature shifts $\Delta\beta$ in Fig.~\ref{fig-5:dbeta_vs_theta_m_vs_theta_c} are smaller than those in 
Fig.~\ref{fig-3:dbeta_vs_theta_m_vs_theta_c}. These plots confirm that the coherence encoded in the state of the control qubit is the principal resource enabling ICO-induced modifications of thermalization.

\bibliography{references}
\end{document}